\documentclass[cameraready]{Interspeech}

\usepackage{cite}
\usepackage{hyperref}  
\usepackage{amsmath,amssymb,amsfonts}
\allowdisplaybreaks
\usepackage{algorithmic}
\usepackage{graphicx}
\usepackage{subcaption} 
\usepackage{textcomp}
\usepackage{xcolor}
\usepackage{booktabs}
\usepackage{multirow}
\usepackage{array}
\usepackage{makecell}
\usepackage{tikz}

\title{Geometrically Constrained Decentralized Independent Vector Analysis for Distributed~Microphone~Arrays}

\author{
Changda Chen$^{1}$, 
Yichen Yang$^{2}$, 
Wei Liu$^{1,3}$, 
Bing Zhu$^{2}$, \\
Gongping Huang$^{3}$,
Shoji Makino$^{1}$,
Shuai Wang$^{4}$
}

\address{
    $^1$Waseda University, Japan \\
    $^2$Northwestern Polytechnical University, Xi’an, China \\
    $^3$School of Electronic Information, Wuhan University, Wuhan, China \\
    $^4$School of Intelligence Science and Technology, Nanjing University, Suzhou, China
}

\email{changda.chen@toki.waseda.jp}


\usepackage{comment}

\begin{document}

\maketitle

\begin{abstract}
This paper proposes a geometrically constrained decentralized independent vector analysis (GC-Dec-IVA) method for distributed microphone arrays. Recently proposed Dec-IVA method enables source separation by exchanging only power-related statistics to exploit cross-array information. However, this initial attempt often provides negligible improvement over applying IVA locally at each array, mainly due to the potential permutation inconsistency among arrays and the strong cross-array dependency implied by its source model. To address these limitations, we incorporate direction-of-arrival (DOA) information to derive GC-Dec-IVA, which mitigates permutation mismatch across arrays and enhances source alignment. Furthermore, a new source model is introduced to weaken cross-array dependency, improving robustness against permutation inconsistency in noisy environments. Experiments show the proposed method improves both the separation performance and cross-array permutation consistency.
\end{abstract}

\section{Introduction}
Blind source separation (BSS) aims to separate acoustic source signals from the observed mixtures on a microphone array without prior knowledge of the mixing process \cite{makino2018ass}. Independent component analysis (ICA) \cite{comon1994independent,smaragdis1998FD-ICA,hyvarinen2000independent,makino2004audio} is a fundamental BSS method that maximizes the statistical independence among the sources to perform separation. Its extension, independent vector analysis (IVA) \cite{hiroe2006solution,kim2006blind,kim2006independent}, has now become a standard approach for speech source separation in the short-time Fourier transform (STFT) domain. By modeling cross-frequency dependency through source vectors, IVA effectively mitigates the frequency-wise permutation problem that often arises in frequency-domain BSS and has demonstrated strong performance in reverberant environments.

In many applications such as meeting transcription, teleconferencing, and smart spaces, acoustic signals can be captured by distributed microphone arrays, where multiple small arrays are deployed over a wide area such as conference rooms, offices, and other indoor spaces \cite{bertrand2011DMA,bertrand2015DMA,cobos2017DMA}. Although these distributed observations provide richer spatial diversity, they also pose a challenge for BSS in how to exploit information across arrays efficiently. Existing strategies can be broadly grouped into the following categories: (i) locally applying BSS at each array, which is communication-free but cannot leverage spatial information from the other arrays and the permutation of the outputs at different arrays is usually inconsistent; (ii) selecting one or a subset of arrays to process \cite{zhang2021select,gunther2021select,hu2023select,ikeshita2025select}, whose performance depends on the selection strategy that varies in different situations and still fails to exploit distributed observations; and (iii) centralized methods that jointly process signals from all microphones \cite{wang2016centralized}. Such centralized processing of all arrays requires transmitting multichannel signals, which increases communication overhead and raises privacy concerns.

Recently proposed decentralized independent vector analysis (Dec-IVA) \cite{yamaoka2025auxiliary} based on the auxiliary-function technique \cite{ono2011stable}  introduces a new approach that performs IVA by exchanging the power-related statistics rather than microphone signals.
The exchanged information can be seen as the shared source activity, which is aggregated across arrays to build an auxiliary variable from the source model.
This approach offers a new perspective to aggregate information from different arrays and avoids the direct transmission of microphone signals. 
Nevertheless, Dec-IVA often yields negligible separation performance gain over running IVA locally at each array, which we refer to as Loc-IVA. A key limitation lies in the globally shared source activity measure, which implicitly assumes that, for each output index $n$, the $n$-th separated signal at each array corresponds to the same speech source. Due to the permutation-ambiguous nature of BSS, different arrays may
assign different source indices to the same speaker. 
When such mismatch occurs, the shared source
activity measure will mix energies from different sources and misguide the updates. Moreover, the source model in \cite{yamaoka2025auxiliary} assumes a strong cross-array dependency, which may amplify the permutation mismatch, particularly in the presence of background noise.

Motivated by the above issues, in this paper, we propose a geometrically constrained decentralized independent vector analysis (GC-Dec-IVA) method for distributed microphone arrays. We incorporate direction-of-arrival (DOA) information into the original decentralized method to derive a new optimization algorithm. The DOA-based geometric constraint is shown to be effective in the field of source separation \cite{parra2002geometric, brendel2020unified,li2020geometrically,wang2023spatially,yang2023geometrically,chen2025switching,liu2025microphone}. In this work, the DOA information can act as a prior on the demixing filters to encourage all arrays to assign the same speaker to each output index, thereby enforcing cross-array source alignment. In addition, we introduce a new source model that better captures relations between arrays compared with the original decentralized method, improving robustness to cross-array permutation mismatch. Experiments in simulated reverberant rooms with 2 to 8 microphone arrays, under both noiseless and noisy conditions, demonstrate that the proposed method consistently improves both separation performance and cross-array permutation consistency compared with the original decentralized and local methods.

\section{Signal model and Problem formulation}
In an acoustic environment, suppose there are $P$ microphone arrays, each equipped with $M$ microphones, observing mixtures of $N$ speech sources in the STFT domain. 
For simplicity, we assume that all arrays contain the same number of microphones and focus on the determined case $M = N$ in this paper.
Nevertheless, the formulation can be naturally extended to the overdetermined case $M \geq N$.
Following the Dec-IVA \cite{yamaoka2025auxiliary} method, the mixture signals observed at the $p$-th array can be modeled as
\begin{align}
{\mathbf{x}}_{f',t} = {\mathbf{A}}_{f'}\,{\mathbf{s}}_{f',t} 
\label{mixing}, 
\end{align}
where $p\in\{1,\ldots,P\}$, $f\in\{1,\ldots,F\}$, and $t\in\{1,\ldots,T\}$ denote the array, original frequency bin, and time frame indices, respectively, and $f' = f + (p-1)F~\in\{1,\ldots,PF\}$ is an extended frequency index obtained by stacking the $F$ frequency bins of the signals observed or processed at each array. Here, ${\mathbf{A}}_{f'} \in \mathbb{C}^{M\times N}$ is the mixing matrix, and the speech source signals ${\mathbf{s}}_{f',t}$ and observed mixture signals ${\mathbf{x}}_{f',t}$ can be respectively expressed as
\begin{align}
{\mathbf{s}}_{f',t} &= 
[{S}_{1,f',t},~\ldots,~ {S}_{N,f',t}]^{\mathsf{T}}
\in \mathbb{C}^{N\times 1},
\\
{\mathbf{x}}_{f',t}  &= 
[{X}_{1,f',t},~\ldots,~ {X}_{M,f',t}]^{\mathsf{T}}
\in \mathbb{C}^{M\times 1},
\end{align}
where $(\cdot)^\mathsf{T}$ denotes the transpose operator. 

Assuming that ${\mathbf{A}}_{f'}$ is invertible, the corresponding demixing process can be modeled as
\begin{align}
{\mathbf{y}}_{f',t}={\mathbf{W}}_{f'}\,{\mathbf{x}}_{f',t}
\label{demixing},
\end{align}
where the demixing matrix ${\mathbf{W}}_{f'}={\mathbf{A}}_{f'}^{-1}$ and separated source signals ${\mathbf{y}}_{f',t}$ are respectively expressed as
\begin{align}
{\mathbf{W}}_{f'} &= 
[{\mathbf{w}}_{1,f'},~\ldots,~ {\mathbf{w}}_{N,f'}]^{\mathsf{H}}
\in \mathbb{C}^{N\times M},
\\
{\mathbf{y}}_{f',t} &= 
[{Y}_{1,f',t},~\ldots,~ {Y}_{N,f',t}]^{\mathsf{T}}
\in \mathbb{C}^{N\times 1},
\end{align}
where $(\cdot)^\mathsf{H}$ denotes the Hermitian transpose operator. To obtain the separated source signals at each array, we build on the auxiliary-function-based IVA method \cite{ono2011stable} to estimate ${\mathbf{W}}_{f'}$ for $1\leq p \leq P$ in this paper.

For Dec-IVA, the auxiliary function it used can be expressed as \cite{yamaoka2025auxiliary}
\begin{align}
Q_{\text{Dec-IVA}}(\mathcal{W},\mathcal{V})
&= 
\sum_{f'=1}^{PF}
\Big(
\sum_{n=1}^{N}
{\mathbf{w}}_{n,f'}^{\mathsf{H}}
{\mathbf{V}}_{n,f'}
{\mathbf{w}}_{n,f'}\nonumber
\\
&\quad - 2\log\left|\det {\mathbf{W}}_{f'}\right|
  +\kappa \Big)
  \label{Dec-IVA},
\end{align}
where $\mathcal{W}=\{{\mathbf{W}}_{f'}\}_{f'}$,  $\mathcal{V}=\{{\mathbf{V}}_{n,f'}\}_{n,f'}$, $n\in\{1,\ldots,N\}$ denotes the index of the separated source signal, and $\kappa$ denotes a constant term independent of $\mathbf{W}_{f'}$. Here, ${\mathbf{V}}_{n,f'}$ serves as the auxiliary variable and can be seen as the weighted covariance for the $n$-th source, expressed as 
\begin{align}
 {\mathbf{V}}_{n,f'}=\frac{1}{T}\sum_{t=1}^T \varphi (r_{n,t}){\mathbf{x}}_{f',t}{\mathbf{x}}_{f',t}^{\mathsf{H}} \label{weighted covariance},
\end{align}
where
\begin{align}
   r_{n,t}=\sqrt{\sum_{f'=1}^{PF}\lvert \mathbf{w}_{n,f'}^{\mathsf{H}}\mathbf{x}_{f',t}\rvert^2}, ~~~~
   \varphi (r_{n,t}) = \frac{G'_{R}(r_{n,t})}{2r_{n,t}}.
\end{align}
Here, $r_{n,t}$ is computed by aggregating the power of the $n$-th separated source signal over all stacked frequency bins, $G_{R}(r_{n,t})$ is the so-called contrast function that follows a probability density function of the source prior, and  $\varphi (r_{n,t})$ is derived to represent the source model that builds the cross-frequency and cross-array relations \cite{ono2011stable}. In the original Dec-IVA \cite{yamaoka2025auxiliary}, the source prior follows a spherical Laplace distribution, and the source model can be expressed as 
\begin{equation}
 \varphi (r_{n,t})
 = \frac{1}{r_{n,t}}
= \frac{1}{\sqrt{\sum_{f'=1}^{PF}\lvert \mathbf{w}_{n,f'}^{\mathsf{H}}\mathbf{x}_{f',t}\rvert^2}}
\label{source model 1}.
\end{equation}
Therefore, at each array, Dec-IVA estimates the demixing matrix by exploiting power-related statistics shared among all arrays.
The shared source activity measure $r_{n,t}$ in \eqref{source model 1}
globally couples all stacked frequency bins, and implicitly assumes that the $n$-th output
${\mathbf{w}}_{n,f'}^{\mathsf{H}}{\mathbf{x}}_{f',t}$ 
corresponds to the same speech source at each array.
However, due to the permutation ambiguity inherent in BSS,
different arrays may assign different source indices to the same speaker.
Consequently, although this modeling leverages the shared source activation assumption to partially mitigate cross-array permutation inconsistency,
$r_{n,t}$ may incorrectly aggregate energies from different sources,
misguiding the parameter updates and leading to block-permutation-like failures.
Moreover, because frequency components observed at different arrays are usually weakly dependent in practice, the uniform cross-array dependency structure imposed by the source model can amplify such misalignment errors, especially in noisy environments.


This motivates us to introduce a DOA-informed geometric constraint to enforce cross-array source alignment and redesign the source model to weaken cross-array dependency induced by the original source model.

\section{Proposed GC-Dec-IVA method}
\subsection{Geometrically constrained cost function}
To incorporate prior DOA information into the optimization of Dec-IVA, we adopt the maximum a posteriori (MAP) principle to derive the cost function \cite{brendel2020unified}. Based on the auxiliary function \eqref{Dec-IVA}, we derive the MAP-based cost function of GC-Dec-IVA, given by 
\begin{align}
    \mathcal{J}_{\text{GC-Dec-IVA}}(\mathcal{W},\mathcal{V})=\mathcal{Q}_{\text{Dec-IVA}}(\mathcal{W},\mathcal{V}) - \log p(\mathcal{W})\label{GC-Dec-IVA},
\end{align}
where $ p(\mathcal{W})$ represents the prior distribution of the demixing matrix. Since we want to restrict the source separated at each array to correspond to the same speaker, the prior related to the DOA information can be described as 
\begin{equation}
\begin{aligned}
\log p(\mathcal{W})
=
-\sum_{p=1}^{P}
\sum_{f'=(p-1)F+1}^{pF}
&\sum_{n=1}^{N}
\sum_{i=1}^{N}
\lambda_{p,n,i}
\\
&\quad \cdot
\bigl|
\mathbf{w}_{n,f'}^{\mathsf{H}}\mathbf{d}_{f',\theta_{p,i}} - c_{n,i}
\bigr|^{2},\label{eq:doa_prior}
\end{aligned}
\end{equation}
where $\theta_{p,i}\in\Theta_p$, $\Theta_p=\{\theta_{p,1},\ldots,\theta_{p,N}\}$ is the set of DOAs of the $N$ sources observed at the $p$-th array, ${\mathbf{d}}_{f',\theta_{p,i}}$ is the corresponding steering vector that follows the same stacking principle as in \eqref{mixing}, $i$ denotes the index that enumerates the candidate DOAs, and $\lambda_{p,n,i}$ is the weight of the constraint.
At the $p$-th array, this constraint enforces the directional response $\mathbf{w}_{n,f'}^{\mathsf{H}}\mathbf{d}_{f',\theta_{p,i}}$ of the $n$-th demixing vector ${\mathbf{w}}_{n,f'}$ at the DOA $\theta_{p,i}$ to match a non-negative value $c_{n,i}$. In particular, a value of $c_{n,i}\geq1$ indicates signal enhancement from the desired target direction, while a smaller value corresponds to suppression toward an interfering direction.
By adjusting this constraint, all arrays are encouraged to preserve the same target source or suppress the same interferers for the $n$-th output, making the $n$-th output more likely to correspond to the same speaker at each array and thereby promoting cross-array permutation consistency.

\subsection{Optimization algorithm and Proposed source model}
 Since \eqref{GC-Dec-IVA} does not have a closed-form solution, the vectorwise coordinate descent (VCD) algorithm \cite{wright2015coordinate,mitsui2018vectorwise} is used to minimize it, and the parameter sets $\mathcal{V}$ and $\mathcal{W}$ are iteratively updated. 
 
 \subsubsection{Update of $\mathcal{V}$}
To weaken cross-array dependency, inspired by the sub-frequency-band modeling in \cite{liang2012overcoming}, we propose a new source model
 \begin{align}
    \varphi (r_{n,t})
    =
\sum_{p=1}^{P}\frac{1}{\sqrt{\sum_{f'=(p-1)F+1}^{pF}|{\mathbf{w}}_{n,f'}^{\mathsf{H}}{\mathbf{x}}_{f',t}|^2}} \label{source model 2}.
\end{align}
The weighted covariance is then updated by \eqref{weighted covariance}. For simplicity, we denote the decentralized algorithm using the original source model in \eqref{source model 1} as I, and that using the proposed source model in \eqref{source model 2} as II hereafter. The proposed source model treats the frequency bins corresponding to each array as a sub-frequency band and explicitly splits the source activity measures among different arrays, thereby making the output contribution of each individual array more independent from the others. When one array produces a different permutation order from the others, the proposed source model does not treat the frequency components from different arrays as having the same weighting relationship and penalizes such inconsistency more strongly, whereas under the original source model in \eqref{source model 1}, the resulting numerical error tends to be averaged out, which degrades robustness to additive noise. Note that the new source model does not increase the communication cost of the power-related statistics to be exchanged, and the order of the array is not required.

\subsubsection{Update of $\mathcal{W}$}
Keeping $\mathcal{V}$ fixed, we continue to use VCD to update the $n$-th demixing vector ${\mathbf{w}}_{n,f'}$ at the $p$-th array. The  derivative of \eqref{GC-Dec-IVA} with respect to ${\mathbf{w}}_{n,f'}^\mathsf{H}$ is deduced as 
\begin{align}
\frac{\partial\mathcal{J}_{\text{GC-Dec-IVA}}}{\partial \mathbf{w}_{n,f'}^\mathsf{H}}
     &=
    \mathbf{V}_{n,f'}\mathbf{w}_{n,f'} 
    -
    \frac{\mathbf{W}_{f'}^{-1}\mathbf{e}_n}{\mathbf{w}_{n,f'}^\mathsf{H}\mathbf{W}_{f'}^{-1}\mathbf{e}_n}\nonumber
    \\
    &\quad+
    \sum_{i=1}^N\lambda_{p,n,i}
{\mathbf{d}}_{f',\theta_{p,i}}{\mathbf{d}}_{f',\theta_{p,i}}^{\mathsf{H}}\mathbf{w}_{n,f'} \nonumber
\\
 &\quad-
  \sum_{i=1}^N\lambda_{p,n,i}c_{n,i}
{\mathbf{d}}_{f',\theta_{p,i}}\label{eq:derivative},
\end{align}
where $\mathbf{e}_n$ is the vector representing the $n$-th column of the $M\times M$ unit matrix. By setting \eqref{eq:derivative} to zero, we can obtain the update formula of $\mathbf{w}_{n,f'}$. The solution without detailed derivation is given as follows due to the space limitation,
\begin{align}
\mathbf{D}_{n,f'}
&=
{\mathbf{V}}_{n,f'}+
\sum_{i=1}^{N}
\lambda_{p,n,i}
{\mathbf{d}}_{f',\theta_{p,i}}{\mathbf{d}}_{f',\theta_{p,i}}^{\mathsf{H}}
\\
\mathbf{u}_{n,f'}
&=
\mathbf{D}_{n,f'}^{-1}\mathbf{W}_{f'}^{-1}\mathbf{e}_n
\\
\hat{\mathbf{u}}_{n,f'}
&=
\sum_{i=1}^{N}\lambda_{p,n,i}
c_{n,i}
\mathbf{D}_{n,f'}^{-1}
{\mathbf{d}}_{f',\theta_{p,i}}
\\
h_{n,f'}
&=
\mathbf{u}_{n,f'}^{\mathsf{H}}\mathbf{D}_{n,f'}\mathbf{u}_{n,f'}
\\
\hat{h}_{n,f'}
&=
\hat{\mathbf{u}}_{n,f'}^{\mathsf{H}}\mathbf{D}_{n,f'}\mathbf{u}_{n,f'}
\\
\mathbf{w}_{n,f'}
&=
\begin{cases}
\frac{1}{\sqrt{h_{n,f'}}}\mathbf{u}_{n,f'}+\hat{\mathbf{u}}_{n,f'}, 
~~~~~~~~~~~~~~~~~~~~~~~\text{if}~\hat{h}_{n,f'}=0,
\\[2pt]
\frac{\hat{h}_{n,f'}}{2h_{n,f'}}
\left[-1\!+\sqrt{1+\frac{4h_{n,f'}}{|\hat{h}_{n,f'}|^2}}\right]\mathbf{u}_{n,f'}
+\hat{\mathbf{u}}_{n,f'}, 
\text{else}.
\end{cases}
\end{align}

After iteratively updating $\mathcal{V}$ and $\mathcal{W}$ for a fixed number of iterations, the separated source signals can be obtained via \eqref{demixing}.

\begin{table*}[t]
\centering

\caption{Average SDRi (dB) and SIRi (dB) under noiseless (SNR $=\infty$) and noisy (SNR uniformly sampled from $[15,25]$~dB) conditions.}
\label{tab:sdr_sir_main}

\scriptsize
\setlength{\tabcolsep}{2.2pt}
\renewcommand{\arraystretch}{1.05}

\newcommand{\wMethod}{0.16\textwidth}
\newcommand{\wNum}{0.04\textwidth} 

\begin{tabular}{
>{\raggedright\arraybackslash}m{\wMethod}
!{\vrule width 0.8pt}
>{\centering\arraybackslash}p{\wNum}
>{\centering\arraybackslash}p{\wNum}
|>{\centering\arraybackslash}p{\wNum}
>{\centering\arraybackslash}p{\wNum}
|>{\centering\arraybackslash}p{\wNum}
>{\centering\arraybackslash}p{\wNum}
|>{\centering\arraybackslash}p{\wNum}
>{\centering\arraybackslash}p{\wNum}
!{\vrule width 1.2pt}
>{\centering\arraybackslash}p{\wNum}
>{\centering\arraybackslash}p{\wNum}
|>{\centering\arraybackslash}p{\wNum}
>{\centering\arraybackslash}p{\wNum}
|>{\centering\arraybackslash}p{\wNum}
>{\centering\arraybackslash}p{\wNum}
|>{\centering\arraybackslash}p{\wNum}
>{\centering\arraybackslash}p{\wNum}
}
\toprule
\multirow{3}{*}{Method}
& \multicolumn{8}{c!{\vrule width 1.5pt}}{Noiseless condition}
& \multicolumn{8}{c}{Noisy condition} \\
\cmidrule(lr){2-9}\cmidrule(lr){10-17}
& \multicolumn{2}{c|}{2 arrays}
& \multicolumn{2}{c|}{4 arrays}
& \multicolumn{2}{c|}{6 arrays}
& \multicolumn{2}{c!{\vrule width 1.5pt}}{8 arrays}
& \multicolumn{2}{c|}{2 arrays}
& \multicolumn{2}{c|}{4 arrays}
& \multicolumn{2}{c|}{6 arrays}
& \multicolumn{2}{c}{8 arrays} \\
\cmidrule(lr){2-3}\cmidrule(lr){4-5}\cmidrule(lr){6-7}\cmidrule(lr){8-9}
\cmidrule(lr){10-11}\cmidrule(lr){12-13}\cmidrule(lr){14-15}\cmidrule(lr){16-17}
& SDRi & SIRi & SDRi & SIRi & SDRi & SIRi & SDRi & SIRi
& SDRi & SIRi & SDRi & SIRi & SDRi & SIRi & SDRi & SIRi \\
\midrule
Loc-IVA \cite{ono2011stable}& 3.98 & 9.17 & 3.96 & 9.11 & 4.02 & 9.15 & 4.10 & 9.27
& 2.61 & 6.92 & 2.51 & 6.78 & 2.46 & 6.70 & 2.44 & 6.70 \\
Dec-IVA I \cite{yamaoka2025auxiliary}
& 4.05 & 9.22 & 4.00 & 9.14 & 4.24 & 9.48 & 4.32 & 9.60
& 2.04 & 6.06 & 1.18 & 4.93 & 0.80 & 4.38 & 0.29 & 3.74 \\
\makecell[tl]{Dec-IVA II (prop.)}& 4.20 & 9.45 & 4.22 & 9.54 & 4.35 & 9.67 & 4.54 & 9.99
& 2.85 & 7.26 & 2.45 & 6.80 & 2.25 & 6.50 & 2.35 & 6.69 \\
GC-Loc-IVA \cite{li2020geometrically}
& 4.56 & 10.09 & 4.56 & 10.04 & 4.53 & 9.97 & 4.58 & 10.03
& 3.21 & 7.99 & 3.19 & 7.94 & 3.03 & 7.69 & 2.97 & 7.61 \\
\makecell[tl]{GC-Dec-IVA I (prop.)}
& 4.56 & 10.01 & 4.69 & 10.23 & \textbf{4.81} & \textbf{10.39} & \textbf{4.90} & \textbf{10.53}
& 2.85 & 7.45 & 2.32 & 6.61 & 2.02 & 6.13 & 1.67 & 5.64 \\
\makecell[tl]{GC-Dec-IVA II (prop.)}
& \textbf{4.65} & \textbf{10.21} & \textbf{4.74} & \textbf{10.35} & 4.77 & 10.36 & 4.87 & 10.50
& \textbf{3.32} & \textbf{8.17} & \textbf{3.41} & \textbf{8.30} & \textbf{3.37} & \textbf{8.22} & \textbf{3.34} & \textbf{8.18} \\
\bottomrule
\end{tabular}
 \vspace{-8pt}
\end{table*}

\section{Experiments}
\subsection{Experimental setup}

\begin{figure}[t]
  \centering
  \includegraphics[width=0.4\textwidth]{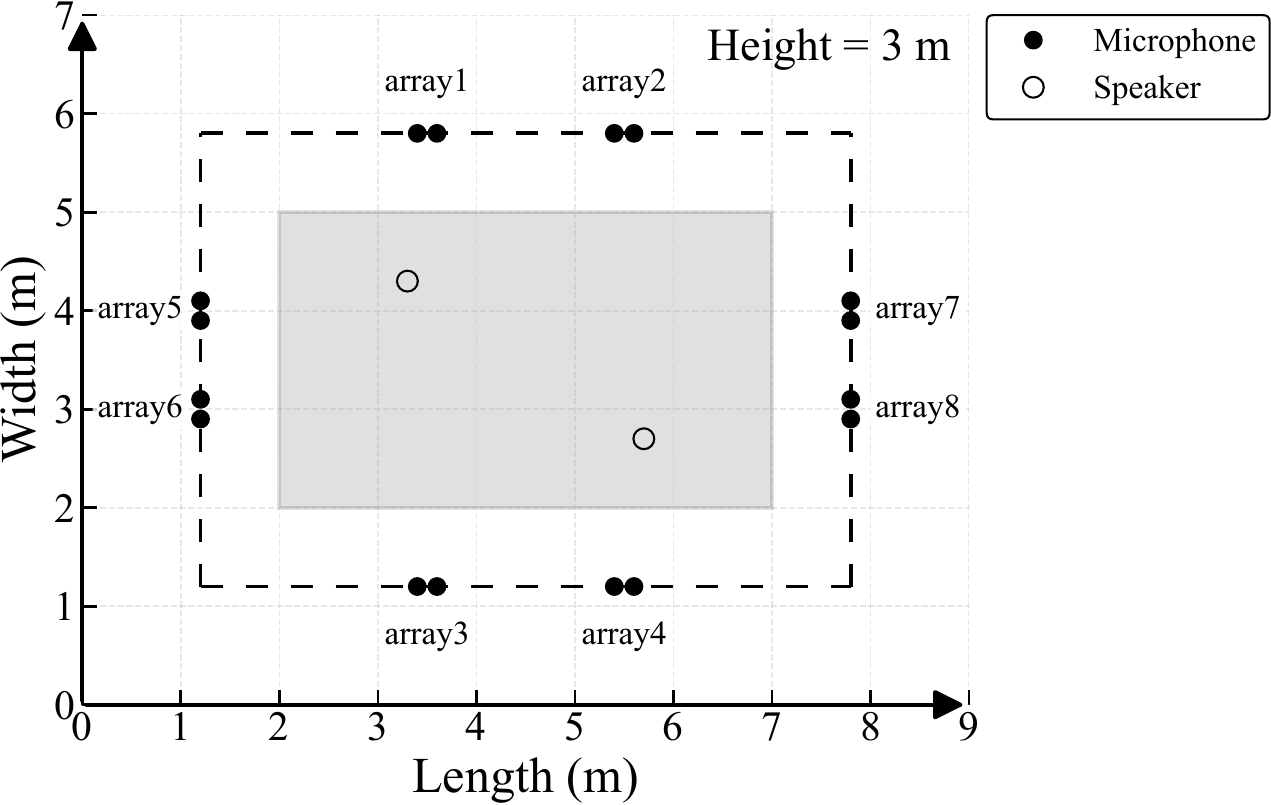}  
  \caption{Simulation layout for signal mixing.
The eight two-microphone arrays are placed at fixed positions with their absolute center coordinates being: 
array1 (3.5, 5.8), array2 (5.5, 5.8),
array3 (3.5, 1.2), array4 (5.5, 1.2),
array5 (1.2, 4.0), array6 (1.2, 3.0),
array7 (7.8, 4.0), and array8 (7.8, 3.0).
The microphone spacing of all arrays is 4 cm. The microphones and speech sources are located at a height of 1.5~m}
  \label{fig:layout}
   \vspace{-10pt}
\end{figure}

We generate 100 ten-second speech mixtures of two speakers sampled at 16~kHz using utterances from the CMU ARCTIC corpus \cite{kominek2004cmu}. The male speaker \textit{rms} and the female speaker \textit{clb} act as the speech sources to be mixed, with their dry signals scaled to have equal energy. To simulate noisy environments, additive background noise consists of diffuse noise \cite{habets2008diffuse} and white Gaussian noise. The diffuse-to-white power ratio is uniformly sampled from $[15,25]$~dB. The total noise power is then scaled to achieve an SNR uniformly sampled from $[15,25]$~dB relative to the reverberant male speech source at each array. The detailed simulation layout is shown in Fig.~\ref{fig:layout}, 
where the speakers are randomly placed within the central grey region, 
with a minimum distance of 1.5~m maintained between the two speakers. We assume that all arrays are synchronized, i.e., no sampling-rate offsets are present. Room impulse responses are generated using the image method \cite{allen1979image} in a $9~\text{m}\times7~\text{m}\times3~\text{m}$ room with the reverberation time $T_{60}=200$~ms. Signals in the STFT domain are computed using a 2048-sample Hann window and a 1024-sample hop size. 

We compare baselines (Loc-IVA \cite{ono2011stable}, Dec-IVA I \cite{yamaoka2025auxiliary}, and GC-Loc-IVA \cite{li2020geometrically}, which applies DOA-based geometric constraints at each array without cross-array information exchange) with the proposed variants (Dec-IVA II, GC-Dec-IVA I, and GC-Dec-IVA II). We adopt the null constraint in \eqref{eq:doa_prior} for geometrically constrained methods. For the $n$-th demixing filter $\mathbf{w}_{n,f'}$, we set $c_{n,i}=0$ for $i\neq n$ towards the interfering source directions, enforcing spatial nulls, while imposing no additional constraint on the target direction. We assume that source DOAs are available at each array. 
In practice, when the relative array geometry is known, 
the DOAs at arrays without direct estimates can be inferred 
from those measured at other arrays. The weight $\lambda_{p,n,i}$ is initialized to 8000 for any $p$, $n$ and $i$, and exponentially decays with factor $0.8$ per iteration following the same rule in \cite{mo2023GC}.
All methods are run for 50 iterations, and projection back is used to resolve the scale ambiguity \cite{murata2001proj-back}.

\subsection{Results and discussions}
We first report the average signal-to-distortion ratio improvement (SDRi) and signal-to-interference ratio improvement (SIRi) \cite{vincent2006performance} over all arrays under both noiseless and noisy conditions, using the reverberant clean source signals of two speech sources at the first microphone of each array as references. Table~\ref{tab:sdr_sir_main} shows that, in the noiseless case, decentralized methods with the original source model in \eqref{source model 1} achieve reasonable performance. However, their performance degrades drastically in the noisy case, and the degradation becomes more pronounced as the number of arrays increases. This indicates that the globally shared source activity coupling imposed by \eqref{source model 1} is sensitive to background noise and has difficulty in modeling cross-array relations in practice.
By contrast, the proposed source model shows robustness in both conditions: Dec-IVA II consistently outperforms Dec-IVA I and yields performance comparable to Loc-IVA in the noisy case. Moreover, the geometrically constrained method with the newly proposed source model GC-Dec-IVA II achieves the best performance under noisy conditions, highlighting the benefit of DOA-guided source alignment for decentralized separation. Interestingly, GC-Dec-IVA I does not outperform GC-Loc-IVA, which further underscores the importance of the proposed source model.

\begin{figure}[t]
  \centering
  \begin{subfigure}[b]{0.475\linewidth}
    \centering
    \includegraphics[width=\linewidth]{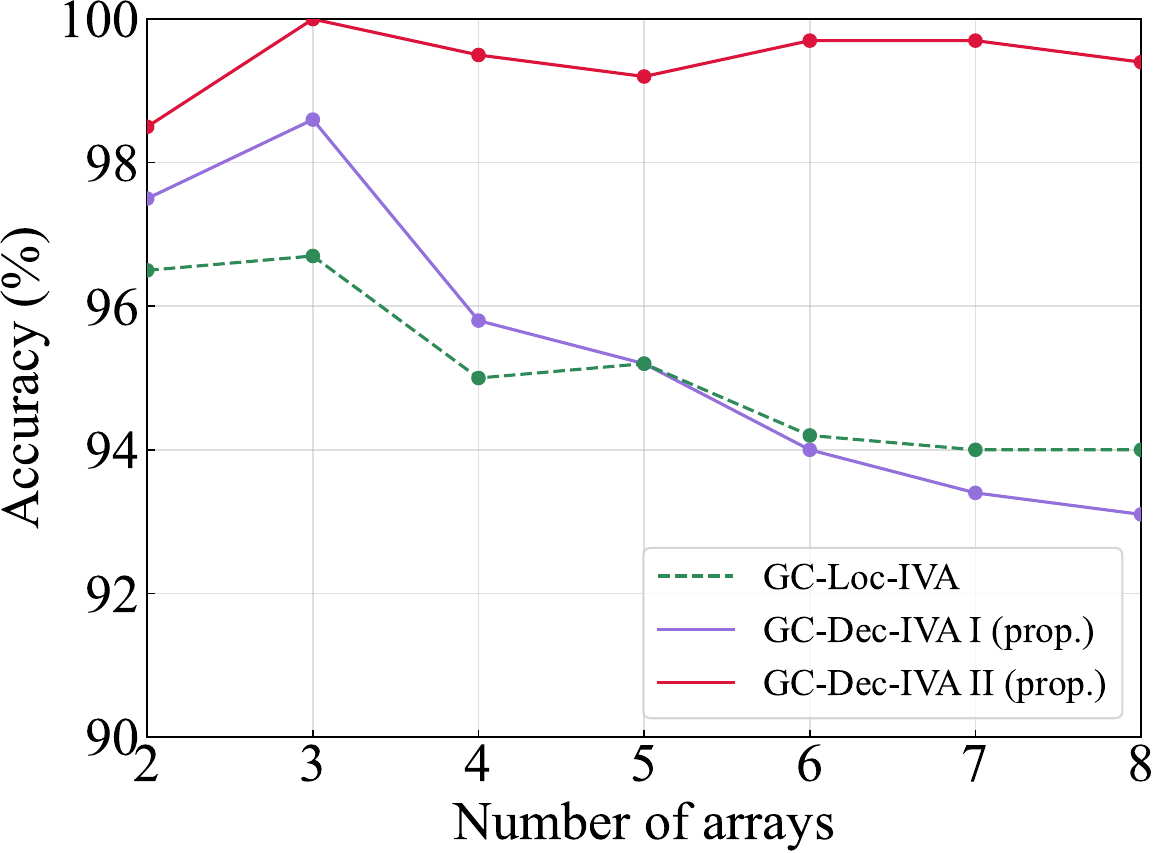}
    \caption{Permutation accuracy (\%)}
    \label{fig:Acc}
  \end{subfigure}
 \hspace{0.03\linewidth}
  \begin{subfigure}[b]{0.475\linewidth} \hskip -20pt
    \centering
    \includegraphics[width=\linewidth]{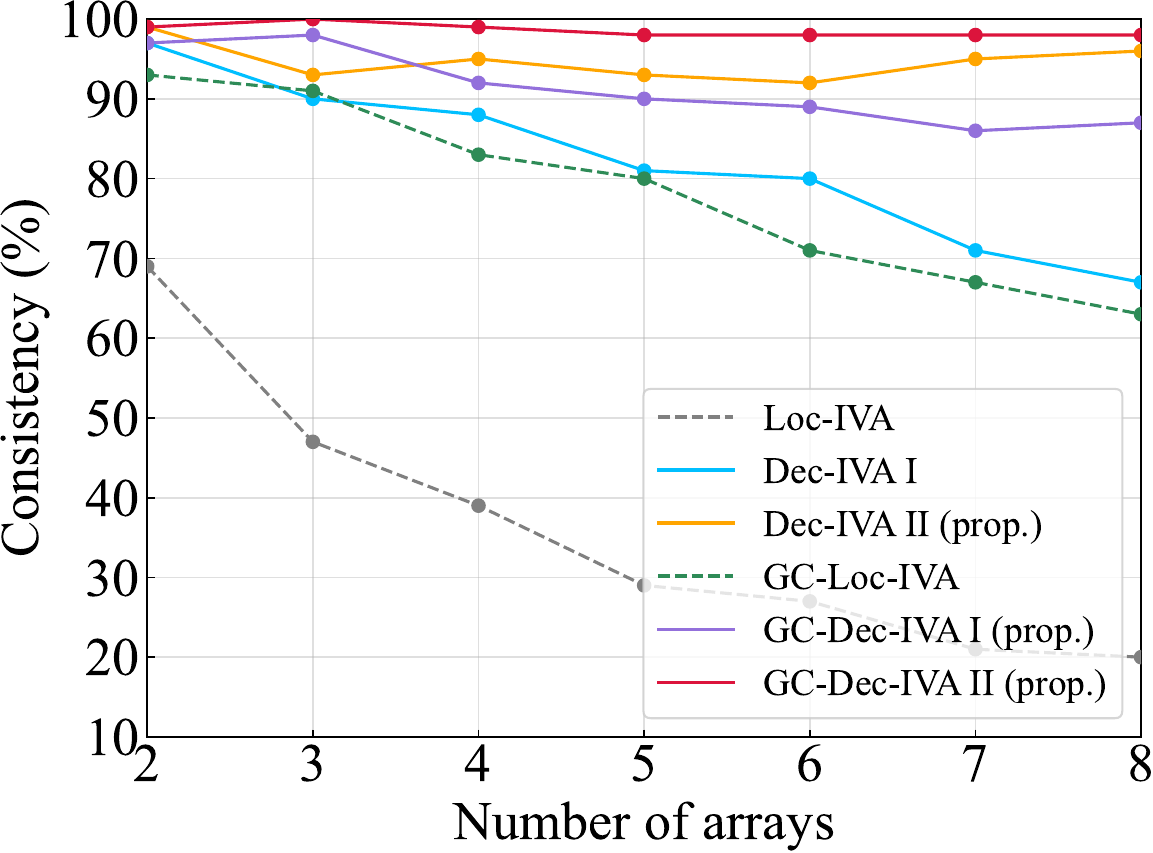}
    \caption{Permutation consistency (\%)}
    \label{fig:Consist}
  \end{subfigure}
  \caption{Permutation accuracy and permutation consistency across 2 to 8 arrays.}
  \label{fig:combined}
 \vspace{-10pt}
\end{figure}

We adopt the noisy condition for subsequent experiments. The average permutation accuracy of the geometrically constrained methods and permutation consistency of all methods across arrays are respectively illustrated in Fig.~\ref{fig:combined}.
Here, the permutation accuracy is computed as the average, over all arrays, of the proportion of mixtures that are separated with the correct permutation, where the correct permutation is defined as the ordering of estimated sources that maximizes the SIR with respect to the reference signals in their original order. 
The permutation consistency is defined as the percentage of cases for which all arrays share an identical permutation, regardless of whether that permutation is correct.
As the number of arrays increases, the local methods intuitively exhibit a rapid drop in permutation consistency because the separation at different arrays is carried out independently without cross-array information exchange during the update process. Additionally, both the accuracy and consistency of the decentralized baselines decrease, reflecting severe cross-array permutation mismatch.
In contrast, the proposed methods substantially improve both metrics: GC-Dec-IVA II achieves near-perfect accuracy and consistency, while Dec-IVA II consistently outperforms its counterpart Dec-IVA I. These results strongly confirm the effectiveness of the proposed source model and DOA-guided geometric constraint.


\begin{table}[t]
\centering

\caption{Performance on arrays that lack DOA information.}
\label{tab:missing_doa}

\scriptsize
\setlength{\tabcolsep}{2.0pt}
\renewcommand{\arraystretch}{1.05}

\newcommand{\wMethod}{0.30\columnwidth}
\newcommand{\wMetric}{0.085\columnwidth} 

\begin{tabular}{
>{\raggedright\arraybackslash}m{\wMethod}
!{\vrule width 0.8pt}
>{\centering\arraybackslash}p{\wMetric}
>{\centering\arraybackslash}p{\wMetric}
>{\centering\arraybackslash}p{\wMetric}
!{\vrule width 1.5pt}
>{\centering\arraybackslash}p{\wMetric}
>{\centering\arraybackslash}p{\wMetric}
>{\centering\arraybackslash}p{\wMetric}}
\toprule
\multirow{2}{*}{Method}
& \multicolumn{3}{c!{\vrule width 1.2pt}}{\makecell[c]{4 arrays\\(array3--4 lack DOA info)}}
& \multicolumn{3}{c}{\makecell[c]{8 arrays\\(array7--8 lack DOA info)}} \\
\cmidrule(lr){2-4}\cmidrule(lr){5-7}
& SDRi & SIRi & Acc.& SDRi & SIRi & Acc.\\
\midrule
\makecell[tl]{Loc-IVA \cite{ono2011stable}}
& 2.42 & 6.65 & 71.50\%
& 2.40 & 6.67 & 79.00\% \\
\makecell[tl]{GC-Dec-IVA I (prop.)}& 1.45 & 5.36 & 90.50\%
& 0.69 & 5.67 & 90.50\% \\
\makecell[tl]{GC-Dec-IVA II (prop.)}& \textbf{2.79} & \textbf{7.31} & \textbf{95.50\%}
& \textbf{2.82} & \textbf{7.38} & \textbf{99.00\%} \\
\bottomrule
\end{tabular}
 \vspace{-10pt}
\end{table}

Finally, we evaluate cases where some arrays lack DOA information by setting the corresponding weight in \eqref{eq:doa_prior} to 0, and report the average performance on these arrays in Table~\ref{tab:missing_doa}. The proposed GC-Dec-IVA II can still recover the correct permutation by leveraging shared information from DOA-informed arrays, while GC-Dec-IVA I exhibits a much worse performance, mainly due to the vulnerability of the original source model in this partially informed setting.

\section{Conclusions}
In this paper, we propose GC-Dec-IVA, aiming at exploiting DOA information to mitigate the underlying cross-array source permutation inconsistency problem in the original Dec-IVA.  Besides, a new source model is introduced to further improve robustness to cross-array permutation mismatch. Experiments in simulated reverberant rooms with 2 to 8 microphone arrays under both noiseless and noisy conditions show that the proposed method consistently improves separation performance and cross-array permutation consistency over the original decentralized and local methods, with particularly pronounced gains in noisy scenarios.


\newpage

\section{Generative AI Use Disclosure}
Generative AI tools were used only for language editing and polishing of the manuscript. The authors have reviewed and verified all AI-assisted edits and take full responsibility for the content, results, and conclusions of this paper. No generative AI tool was used to produce a significant part of this work.

\bibliographystyle{IEEEtran}
\bibliography{mybib}

@book{makino2018ass,
  title={Audio Source Separation},
  author={Makino, Shoji},
  year={2018},
  address={Switzerland},
  publisher={Springer}
}

@article{comon1994independent,
  title={Independent component analysis, a new concept?},
  author={Comon, Pierre},
  journal={Signal Process.},
  volume={36},
  number={3},
  pages={287--314},
  year={Apr. 1994},
  publisher={Elsevier}
}

@article{hyvarinen2000independent,
  title={Independent component analysis: algorithms and applications},
  author={Hyv{\"a}rinen, Aapo and Oja, Erkki},
  journal={Neural Netw.},
  volume={13},
  number={4-5},
  pages={411--430},
  year={Jun, 2000},
  publisher={Elsevier}
}

@inproceedings{makino2004audio,
  title={Audio source separation based on independent component analysis},
  author={Makino, Shoji and Araki, Shoko and Mukai, Ryo and Sawada, Hiroshi},
  booktitle={Proc. ISCAS},
  volume={5},
  pages={V-668--V-671},
  year={2004},
}

@article{smaragdis1998FD-ICA,
  title={Blind separation of convolved mixtures in the frequency domain},
  author={Smaragdis, Paris},
  journal={Neurocomput.},
  volume={22},
  number={1-3},
  pages={21--34},
  year={Nov. 1998},
  publisher={Elsevier}
}

@inproceedings{kim2006independent,
  title={Independent vector analysis: An extension of {ICA} to multivariate components},
  author={Kim, Taesu and Eltoft, Torbj{\o}rn and Lee, Te-Won},
  booktitle={Proc. ICA},
  pages={165--172},
  year={2006},
}

@article{kim2006blind,
  title={Blind source separation exploiting higher-order frequency dependencies},
  author={Kim, Taesu and Attias, Hagai T and Lee, Soo-Young and Lee, Te-Won},
  journal={IEEE/ACM Trans. Audio, Speech, Lang. Process.},
  volume={15},
  number={1},
  pages={70--79},
  year={Dec. 2006},
  publisher={IEEE}
}

@inproceedings{hiroe2006solution,
  title={Solution of permutation problem in frequency domain {ICA}, using multivariate probability density functions},
  author={Hiroe, Atsuo},
  booktitle={Proc. ICA},
  pages={601--608},
  year={2006},
}

@inproceedings{ono2011stable,
  title={Stable and fast update rules for independent vector analysis based on auxiliary function technique},
  author={Ono, Nobutaka},
  booktitle={Proc. WASPAA},
  pages={189--192},
  year={2011},
}

@inproceedings{bertrand2011DMA,
  title={Applications and trends in wireless acoustic sensor networks: A signal processing perspective},
  author={Bertrand, Alexander},
  booktitle={Proc. SCVT},
  pages={1--6},
  year={2011},
}

@article{bertrand2015DMA,
  title={Special issue on wireless acoustic sensor networks and ad hoc microphone arrays},
  author={Bertrand, Alexander and Doclo, Simon and Gannot, Sharon and Ono, Nobutaka and van Waterschoot, Toon},
  journal={Signal Process.},
  volume={107},
  number={C},
  pages={1--3},
  year={Feb. 2015},
}

@article{cobos2017DMA,
  title={A survey of sound source localization methods in wireless acoustic sensor networks},
  author={Cobos, Maximo and Antonacci, Fabio and Alexandridis, Anastasios and Mouchtaris, Athanasios and Lee, Bowon},
  journal={Wireless Commun. Mobile Comput.},
  year={Aug. 2017},
}

@article{zhang2021select,
  title={Deep ad-hoc beamforming},
  author={Zhang, Xiao-Lei},
  journal={Comput. Speech Lang.},
  volume={68},
  year={Jul. 2021},
  number={101201},
}

@inproceedings{gunther2021select,
  title={Network-aware optimal microphone channel selection in wireless acoustic sensor networks},
  author={Gunther, Michael and Afifi, Haitham and Brendel, Andreas and Karl, Holger and Kellermann, Walter},
  booktitle={Proc. IEEE ICASSP},
  pages={820--824},
  year={2021},
}

@article{hu2023select,
  title={Distributed sensor selection for speech enhancement with acoustic sensor networks},
  author={Hu, De and Si, Qintuya and Liu, Rui and Bao, Feilong},
  journal={IEEE/ACM Trans. Audio Speech Lang. Process.},
  volume={31},
  pages={985--999},
  year={Feb. 2023},
}

@article{ikeshita2025select,
  title={Maximizing Predicted Signal-to-Distortion Ratio: A New Microphone Selection Criterion for Beamforming in Acoustic Sensor Networks},
  author={Ikeshita, Rintaro and Nakatani, Tomohiro and Ochiai, Tsubasa and Araki, Shoko},
  journal={IEEE Trans. Audio Speech Lang, Process.},
  year={May 2025},
}

@article{wang2016centralized,
  title={Correlation maximization-based sampling rate offset estimation for distributed microphone arrays},
  author={Wang, Lin and Doclo, Simon},
  journal={IEEE/ACM Trans. Audio, Speech, Lang. Process.},
  volume={24},
  number={3},
  pages={571--582},
  year={Mar. 2016},

}

@inproceedings{yamaoka2025auxiliary,
  title={Auxiliary-Function-Based Decentralized Independent Vector Analysis for Distributed Microphone Arrays},
  author={Yamaoka, Kouei and Morita, Katsuhiro and Takamune, Norihiro and Saruwatari, Hiroshi},
  booktitle={Proc. APSIPA ASC},
  pages={54--59},
  year={2025},
}

@article{parra2002geometric,
  title={Geometric source separation: Merging convolutive source separation with geometric beamforming},
  author={Parra, Lucas C and Alvino, Christopher V},
  journal={IEEE Trans. Speech, Audio Process.},
  volume={10},
  number={6},
  pages={352--362},
  year={Sep. 2002},
  publisher={IEEE}
}

@inproceedings{li2020geometrically,
  title={Geometrically constrained independent vector analysis for directional speech enhancement},
  author={Li, Li and Koishida, Kazuhito},
  booktitle={Proc. IEEE ICASSP},
  pages={846--850},
  year={2020},
}

@inproceedings{yang2023geometrically,
  title={Geometrically constrained source extraction and dereverberation based on joint optimization},
  author={Yang, Yichen and Wang, Xianrui and Brendel, Andreas and Zhang, Wen and Kellermann, Walter and Chen, Jingdong},
  booktitle={Proc. EUSIPCO},
  pages={41--45},
  year={2023},
}

@inproceedings{wang2023spatially,
  title={Spatially informed independent vector analysis for source extraction based on the convolutive transfer function model},
  author={Wang, Xianrui and Brendel, Andreas and Huang, Gongping and Yang, Yichen and Kellermann, Walter and Chen, Jingdong},
  booktitle={Proc. IEEE ICASSP},
  pages={1--5},
  year={2023},
}

@inproceedings{chen2025switching,
  title={Switching Constant Separating Vector for Moving Source Extraction with Geometric Constraints},
  author={Chen, Changda and Yang, Yichen and Zhao, Yuehao and Makino, Shoji and Chen, Jingdong},
  booktitle={Proc. APSIPA ASC},
  pages={13--18},
  year={2025},
}

@inproceedings{liu2025microphone,
  title={Microphone Array Beamforming for Speech Enhancement Based on Dynamic Mode Decomposition},
  author={Liu, Wei and Huang, Gongping and Liu, Xin and Jin, Jilu and Chen, Jingdong and Benesty, Jacob},
  booktitle={Proc. IEEE ICASSP},
  pages={1--5},
  year={2025},
}

@article{brendel2020unified,
  title={A unified probabilistic view on spatially informed source separation and extraction based on independent vector analysis},
  author={Brendel, Andreas and Haubner, Thomas and Kellermann, Walter},
  journal={IEEE Trans. Signal Process.},
  volume={68},
  pages={3545--3558},
  year={Jun. 2020},
  publisher={IEEE}
}

@article{liang2012overcoming,
  title={Overcoming block permutation problem in frequency domain blind source separation when using AuxIVA algorithm},
  author={Liang, Yanfeng and Naqvi, SM and Chambers, J},
  journal={Electron. Lett.},
  volume={48},
  number={8},
  pages={460--462},
  year={Apr. 2012},
}

@inproceedings{mitsui2018vectorwise,
  title={Vectorwise coordinate descent algorithm for spatially regularized independent low-rank matrix analysis},
  author={Mitsui, Yoshiki and Takamune, Norihiro and Kitamura, Daichi and Saruwatari, Hiroshi and Takahashi, Yu and Kondo, Kazunobu},
  booktitle={Proc. IEEE ICASSP},
  pages={746--750},
  year={2018},
}

@article{wright2015coordinate,
  title={Coordinate descent algorithms},
  author={Wright, Stephen J},
  journal={Math. Program.},
  volume={151},
  number={1},
  pages={3--34},
  year={Jun. 2015},
  publisher={Springer}
}

@inproceedings{kominek2004cmu,
  title={The {CMU} Arctic speech databases},
  author={Kominek, John and Black, Alan W},
  booktitle={Proc. ISCA SSW},
  pages={223--224},
  year={2004}
}

@article{habets2008diffuse,
  title={Generating nonstationary multisensor signals under a spatial coherence constraint},
  author={E. A. P. Habets and I. Cohen and S. Gannot},
  journal={J. Acoust. Soc. Am.},
  volume={124},
  number={5},
  pages={2911--2917},
  year={Nov. 2008},
  publisher={AIP Publishing}
}

@article{allen1979image,
  title={Image method for efficiently simulating small-room acoustics},
  author={Allen, Jont B and Berkley, David A},
  journal = {J. Acoust. Soc. Am.},
  volume={65},
  number={4},
  pages={943--950},
  year={Apr. 1979},
  publisher={Acoustical Society of America}
}

@inproceedings{mo2023GC,
  title={On joint dereverberation and source separation with geometrical constraints and iterative source steering},
  author={Mo, Kaien and Wang, Xianrui and Yang, Yichen and Ueda, Tetsuya and Makino, Shoji and Chen, Jingdong},
  booktitle={Proc. APSIPA ASC},
  pages={1138--1142},
  year={2023},
}

@article{murata2001proj-back,
  title={An approach to blind source separation based on temporal structure of speech signals},
  author={Murata, Noboru and Ikeda, Shiro and Ziehe, Andreas},
  journal={Neurocomput.},
  volume={41},
  number={1-4},
  pages={1--24},
  year={Oct. 2001},
  publisher={Elsevier}
}

@article{vincent2006performance,
  title={Performance measurement in blind audio source separation},
  author={Vincent, Emmanuel and Gribonval, R{\'e}mi and F{\'e}votte, C{\'e}dric},
  journal={IEEE/ACM Trans. Audio, Speech, Lang. Process.},
  volume={14},
  number={4},
  pages={1462--1469},
  year={Jul. 2006},
  publisher={IEEE}
}

\end{document}